\documentclass[final,3p,twocolumn]{elsarticle}
\usepackage{graphicx}
\usepackage{amssymb}
\usepackage{amsmath}

\journal{Physics Letters B}

\begin{document}

\begin{frontmatter}

\title{Dependence of spin induced structural transitions on level density and Neutron emission spectra}
\author{Mamta Aggarwal\corref{cor1}}
\cortext[cor1]{corresponding authors: Mamta Aggarwal, mamta.a4@gmail.com}
%\linebreak
\address{Department of Physics, University of Mumbai, Kalina Campus, Mumbai-400098, India}
\begin{abstract}
The impact of spin induced deformation and shape phase transitions on nuclear level density and consequently on neutron emission spectra of the decay of compound nuclear systems $^{112}$Ru to $^{123}$Cs (N $=$ 68 isotones) is investigated in a microscopic framework of Statistical theory of superfluid nuclei. Our calculations are in good accord with experimental data for evaporation residue of $^{119}$Sb$^∗$ and $^{185}$Re$^∗$ and show a strong correlation between spin induced structural transitions and NLD. We find that the inverse level density parameter 'K' increases with increasing spin for all the systems, but it decreases with a deformation or a shape change that results in the enhancement of level dnesity and emission probability. A  sharp shape phase transition from oblate to uncommon prolate non-collective in well deformed nuclei leads to band crossing  and enhancement of level density which fades away while approaching sphericity at or near shell closure manifesting shell effects. \par
\end{abstract}


%\PACS 23.50.+z, 21.10.-k, 21.10.Dr
\end{frontmatter}
%-------------------------------------------------------------------------------%\section{Introduction}
It is now a well-known fact that the density of quantum mechanical states increase rapidly with excitation energy and the nucleus shifts from discreteness to quasi-continuum and continuum where the statistical concepts, in particular, the nuclear level density (NLD) ~\cite{BETHE,ERIC,BOHR,KATARIA,MTN} which is the number of excited levels around an excitation energy, are crucial for the prediction of various nuclear phenomena, astrophysics ~\cite{WALLA} and nuclear technology. Recent measurements of the evaporation spectra of particles emitted from the highly excited compound nuclear systems in a hot and rotating state have provided ~\cite{KAUSHIK,BALARAM,GOHIL,YKGUP,PROY} some information to understand the interdependence between the nuclear level density (NLD) and the key parameters such as excitation energy, isospin and most importantly the angular momentum, collective and non-collective excitations.  The NLD parameter related to the density of the single particle levels near the Fermi surface is influenced by the shell structure and the shape of the nucleus which in turn are profoundly altered by the excitations ~\cite{MAPL,MACO,MAMHIAS,MI,MAJNP,IGNAT,GOOD,ALHA}. Damping of the shell effects on NLD parameter with excitation energy and fadeout of collective enhancement of NLD with shape transition from deformed to spherical, measured recently ~\cite{PROY,PCROUT,KAUSHPLB}, point towards the influence of shell structure on NLD that requires a comprehensive investigation within a microscopic framework which exactly is the objective of this letter. \par

The angular momentum dependence of NLD parameter with the excitation energy and angular momentum, in particular, is a subject of tremendous interest currently and there have been efforts on the theoretical~\cite{CIVIT,SHLOM,DE,KATA,SHLO,BKAGRA,SSHLO,LANG,KOONIN,PUDDU,MPRL,MAK,MAI} and experimental~\cite{HAGEL,GONIN,FINEMAN,FABRIS,DRCHAKR} fronts but a conclusive viewpoint is still far from reach. A recent measurement~\cite{KAUSHIK} of neutron evaporation energy spectra for the decay of $^{119}$Sb$^∗$ in the excitation energy range of $\approx$ 31$-$43 MeV showed that the inverse level density parameter (k$=$A/a) extracted for different angular momentum regions decreases with increasing angular momentum. Another work~\cite{YKGUP} suggested increasing 'K' with the increasing angular momentum for lower J values but predicted a decreasing trend at higher J values whereas our theoretically derived 'K' values for same nuclei showed increasing 'K' for increasing J~\cite{MAK}. In case of heavier compound nucleus $^{185}$Re$*$ ~\cite{GOHIL}, value of 'K' is found to remain almost constant for different J values whereas value of 'K' decreases with increasing J for $^{97}$Tc~\cite{BALARAM}. All these experimental works lack a proper justification and explanations for the variety of predictions regarding the level density depenedence on angular momentum whose interpretation and proper understanding from a theoretical point of view is absolutely necessary. \par

In view of this we compute deformation, shape, nuclear level density and neutron emission probability for the decay of excited compound nuclear systems $^{119}$Sb$^∗$ and neighbouring N=68 isotones from $^{112}$Ru to $^{123}$Cs, which are described as thermodynamical system of fermions in a microscopic framework of statistical model ~\cite{MTN,MAK} incorporating temperature, collective and non-collective rotation degrees of freedom. Our choice of nuclei includes well deformed as well as magic nuclei which paves a way to check the shell effects in NLD variation ~\cite{PCROUT}, which, in this work, are found quite prevalent at the excitation energies corresponding to temperature T$\approx$ 1.0 $-$ 1.3 MeV. Our estimated inverse level density parameter 'K' and neutron evaporation spectra of $^{119}$Sb$^∗$ residue agree with the experimental data and show the reliability of our theoretical model. Further validation of our model comes from the close agreement between our estimated 'K' for a heavier system $^{185}$Re$^*$ and the experimental data ~\cite{GOHIL}. Our systematic study shows a strong influence of spin induced deformation and shape transitions on NLD ~\cite{KAUSHPLB}, and also explains the predictions of decreasing 'K' with spin in the recent works ~\cite{KAUSHIK,GOHIL,PROY}  to be related to the structural changes seen at those spin values, which is the highlight of this work, not seen before in any other work so far to our knowledge. \par

The nuclear level density and neutron emission spectra for the deexcitation of compound nuclear system in hot and rotating state are obtained by the number of neutrons emitted within an energy interval E$_n$ and E$_n$ +dE$_n$ using the equation~\cite{BLAT}
\begin{equation}
dN(E_n) = C E_n \rho (U_{th}) dE_n,
\end{equation} where $\rho$(U$_{th}$) is the nuclear level density of the residual nuclear system. U$_{th}$ = E$^*$ - E$_{rot}$ -S$_n$ - E$_n$ is the internal excitation energy of the residual nucleus after the neutron emission and E$_n$ is average kinetic energy of the outgoing neutron. E$^*$ is the total excitation energy available to the system due to the reaction process which is shared between various degrees of freedom like rotational energy (E$_{rot}$), neutron separation energy (S$_n$) and E$_n$.  C is the normalisation constant. \par
The nuclear level density  $\rho$(U$_{th}$) at an excitation energy U$_{th}$ is obtained by using the expression~\cite{BLAT,HM}
\begin{equation}
\rho(U_{th}) = {(\hbar^2/2 \Theta)^{3/2} (2I +1) {\sqrt a} exp(2 \sqrt{aU_{th}}) \over 12(U_{th}+T)^2},
%\label{equation}
\end{equation}
where 'a' is NLD parameter 
\begin{equation}
a=S^2/4U_{th} 
\end{equation}
which is related to single particle density at fermi level. $\Theta$ is the rigid-body moment of inertia which is obtained~\cite{HAMA} using
\begin{equation}
\Theta_1 = \hbar^2 I {(dE_{rot}  /dI)}^{-1}
%\label{equation}
\end{equation}
\begin{equation}
\Theta_2 = \hbar^2 {(d^2E_{rot} /dI^2)}^{-1}
%\label{equation}
\end{equation}
Eq. (5) is used only when there is band crossing. \par
To determine the equillibrium deformation and shape of the nucleus, It is common to minimize the appropriate free energy F$=$E$-$TS ~\cite{GOODF}. In our work, we trace F minima with respect to intrinsic shape parameters ($\beta$, $\gamma$) which also describe the orientation of the nucleus with respect to its rotation axis. $\gamma$ values range from -180$^o$ (oblate with symmetry axis parallel to the rotation axis) to -120$^o$ (prolate with symmetry  axis perpendicular to rotation axis) and then to -60$^o$ (oblate collective) to 0$^o$ (prolate non-collective) along with $\beta$ ranging from 0 to 0.4 in steps of 0.01 using the equation  
\begin{eqnarray}
F(Z,N,\beta,\gamma,T,M) = E(Z,N,\beta,\gamma,T,M) \nonumber\\ 
 -  T * S(Z,N,\beta,\gamma,T,M)
\end{eqnarray}
where the energy (E), entropy (S) and the excitation energy (E$^*$) are the functions of particle number, deformation and shape along with the orientation with respect to rotation axis and are computed within the theoretical framework ~\cite{MAPL,MACO,MAK} which involves the statistical theory ~\cite{MTN,MAK} and the Macroscopic-Microscopic approch ~\cite{MASHA,MAPL,GAUPLB} (details of which have been adequately described in our earlier works ~\cite{MAPL,MACO,MAK} hence we are avoiding the detailed description of the formalism here. Readers may refer Refs.~\cite{MAPL,MACO,MAK} for details of the formalism.). \par 
\begin{figure}[!htb]
%\tabcapfont
\centerline{%
\begin{tabular}{c}
\vspace{0.44cm}
\includegraphics[scale=.36]{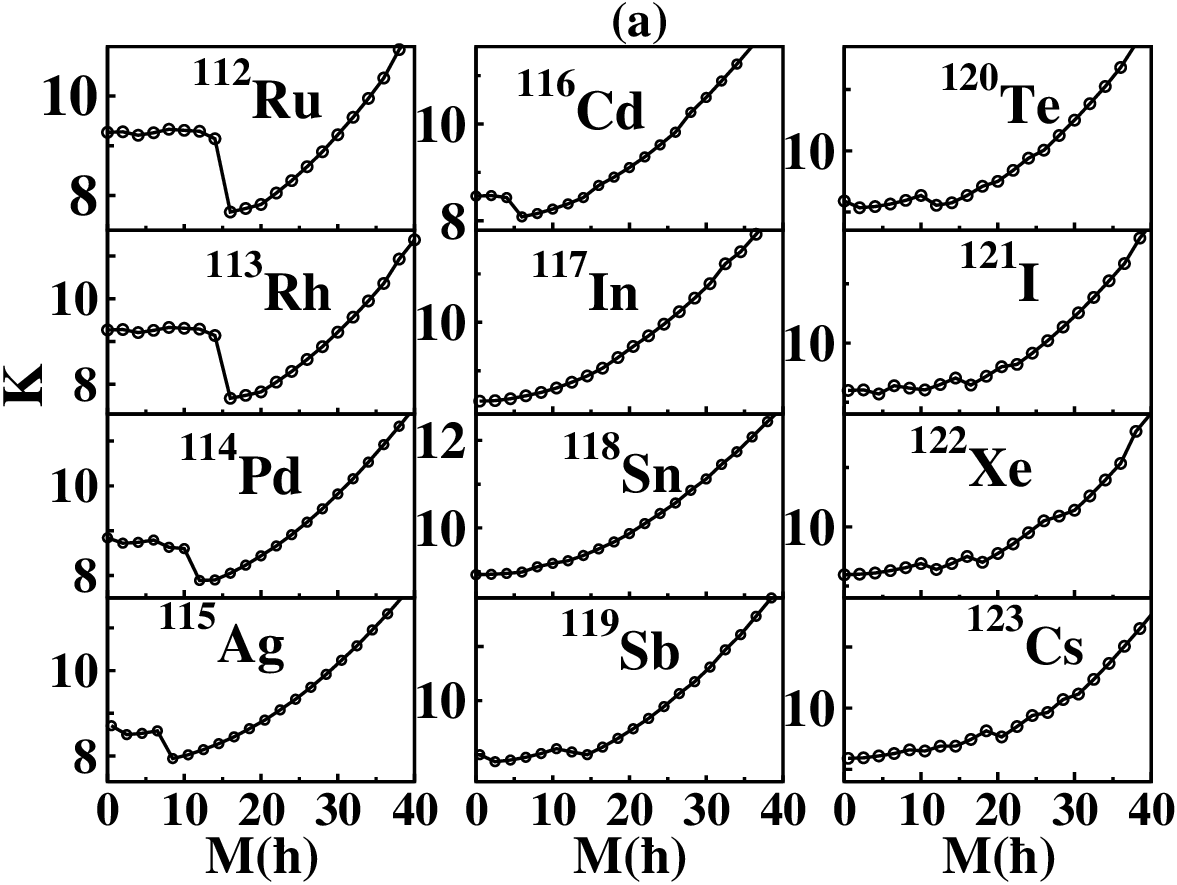} \\
\vspace{0.47cm}
\includegraphics[scale=.36]{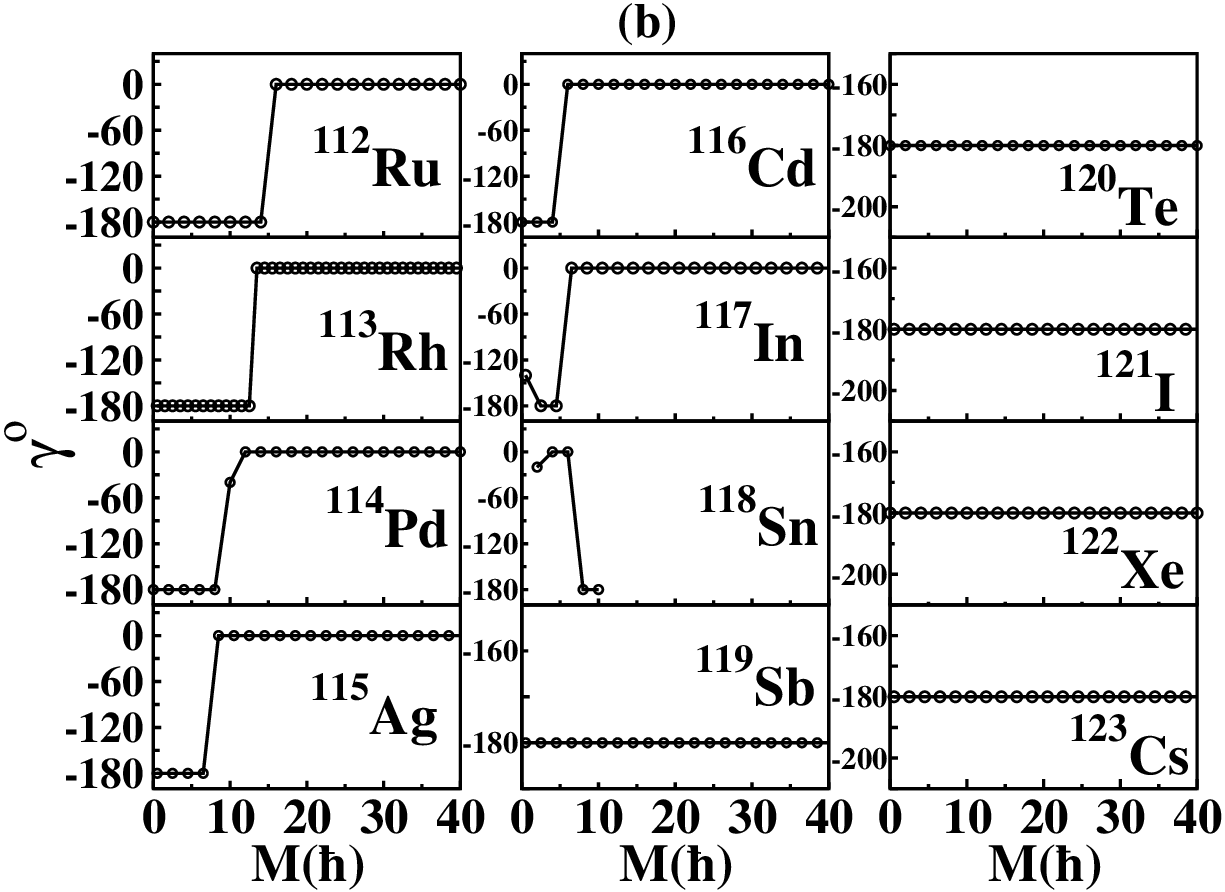}\\
\vspace{0.0cm}
\includegraphics[scale=.36]{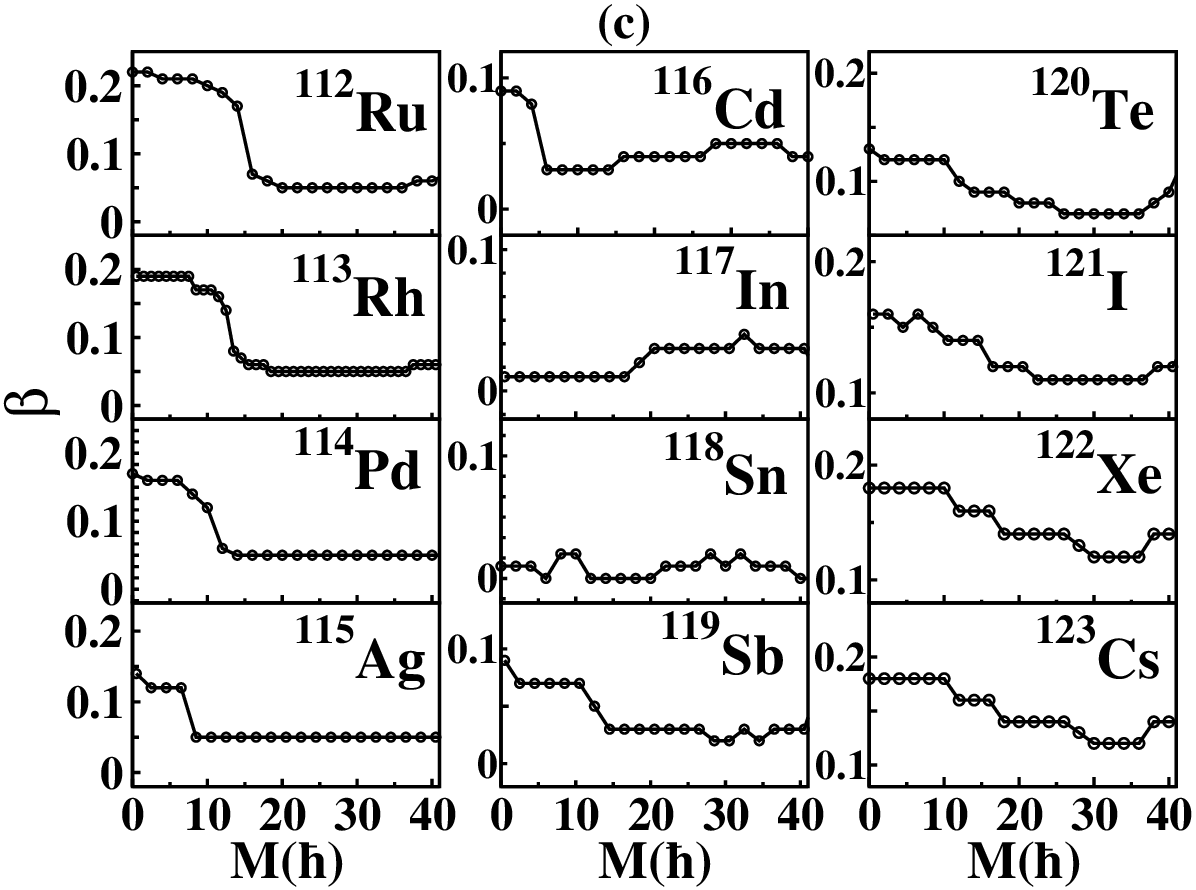}
%a.~~ Octal linked list & b.~~ Quadratic linked list & c.~~Dual linked list
\end{tabular}}
\caption{(a) Inverse level density parameter ’K’ vs. angular momentum M; (b) Shape parameter $\gamma$ vs. M($\hbar$); (c) Equillibrium deformation parameter $\beta$ vs. M($\hbar$).}\label{fig1}
\end{figure}
We evaluate inverse level density parameter K($=$A/a, where 'a' is given by Eq. (3)), shape ($\gamma$) and deformation ($\beta$) as a function of M ranging from $=$0$-$40$\hbar$ at an excitation energy $\approx$ 31 MeV for N=68 isotones from $^{112}$Ru to $^{123}$Cs plotted in Fig. 1(a), 1(b) and 1(c) respectively. We find that 'K' increases with angular momentum for all the nuclear systems studied here (see Fig. 1(a)) which is similar to the observation in our earlier work ~\cite{MAK}, but here, it drops sharply at mid spin values for nuclei $^{112}$Ru, $^{113}$Rh, $^{114}$Pd, $^{115}$Ag $^{116}$Cd where a shape transition (seen in Fig. 1(b)) from oblate ($\gamma$ =-180$^o$) to prolate  non-collective ($\gamma$=0$^o$)  with change in deformation (seen in Fig. 1(c)) takes place. The value of 'K' in the well deformed nucleus $^{112}$ drops significantly from 9.14 MeV at M $=$ 14$\hbar$ (oblate) to 7.67 MeV at  M $=$ 16$\hbar$ (prolate) due to a sharp deformation shape phase transition and where the rotational energy E$_{rot}$ decreases sharply with increasing spin at M $=$ 16$\hbar$ with changing slope in the plot of E$_{rot}$ vs. M  shown in Fig. 2 which indicates band crossing where Eq. (5) has to used for the calculation of moment of inertia. U$_{th}$ increases due to drop in  E$_{rot}$ which leads to enhanced level density and emission probability. These structural transitions and thier impact on 'K', slowly diminish as one moves towards the sphericity at or near shell closure $^{118}$Sn where $\beta$ is very small or almost zero and where $\gamma$ has almost no significance and the shape is mostly spherical manifesting shell effects that influence NLD ~\cite{PCROUT}. The nuclei from $^{119}$Sb to $^{123}$Cs show no shape transitions and remain oblate for all the spin values for which 'K' increases with increasing spin for all M except for few small fluctuations in deformation which are reflected in small fluctuations in 'K'. This establishes a strong correlation between the deformation and shape phase transitions and the  decrease in 'K' with the enhancement of NLD and emission probability in concise with the indication by Refs. ~\cite{KAUSHPLB,PROY}.\par
\begin{figure}
\includegraphics[scale=.36]{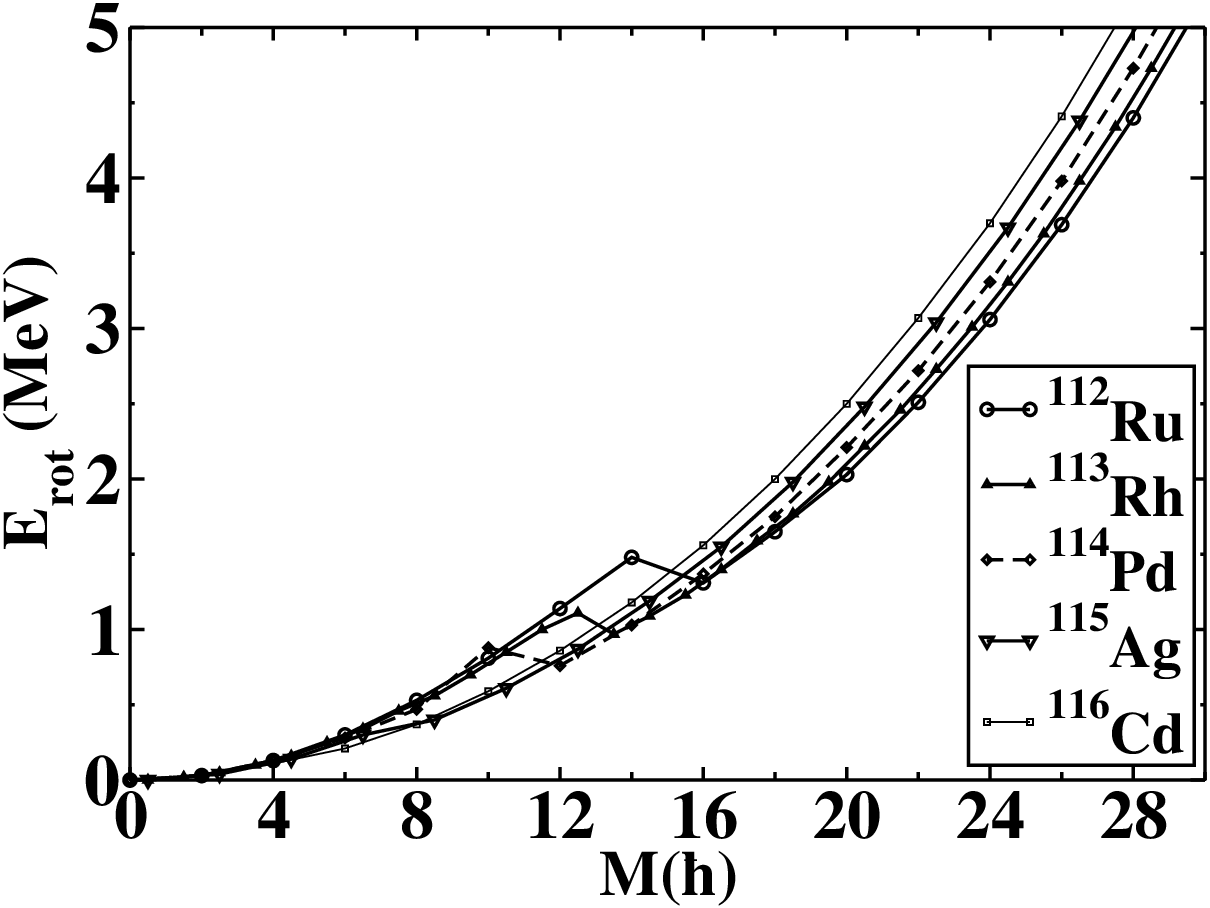}% Here is how to import EPS art
\caption{\label{fig2} Rotational energy E$_{rot}$ vs. M($\hbar$).}
\end{figure}

Here it may be noted that the enhancement of level density along with the band crossing with sudden decline in 'K' value has been observed at a sharp shape phase transition from oblate to prolate non-collective. This rarely seen shape phase has been seen for the first time in neutron rich stable nuclei $^{112}$Ru, $^{113}$Rh, $^{114}$Pd, $^{115}$Ag and $^{116}$Cd that too at  T$>$ 1 MeV. So far we had observed this shape phase only in proton rich nuclei ~\cite{MAPL,MACO} that too for T $<$1 MeV.  However this sharp shape transition may appear considerably blurred if we incorporate the thermal shape fluctuations in the calculations, not done in the present work. Our calculations include spin projections that lead to well defined minima and can considerably reduce the shape fluctuations as pointed out by Refs. ~\cite{HAYASHI,GOODF}. This is evident in our present work where we find well defined F minima defining the deformation, shape and the orientation. Since the changes in the shape and deformation are reflected in the level density parameter variation which has agreed with the experimental data, it shows the efficiency and reliability of our approach. Inspite of certain limitations, our formalism has been able to show the structural transitions and their influence on NLD very efficiently. \par
\begin{figure}
\includegraphics[scale=.36]{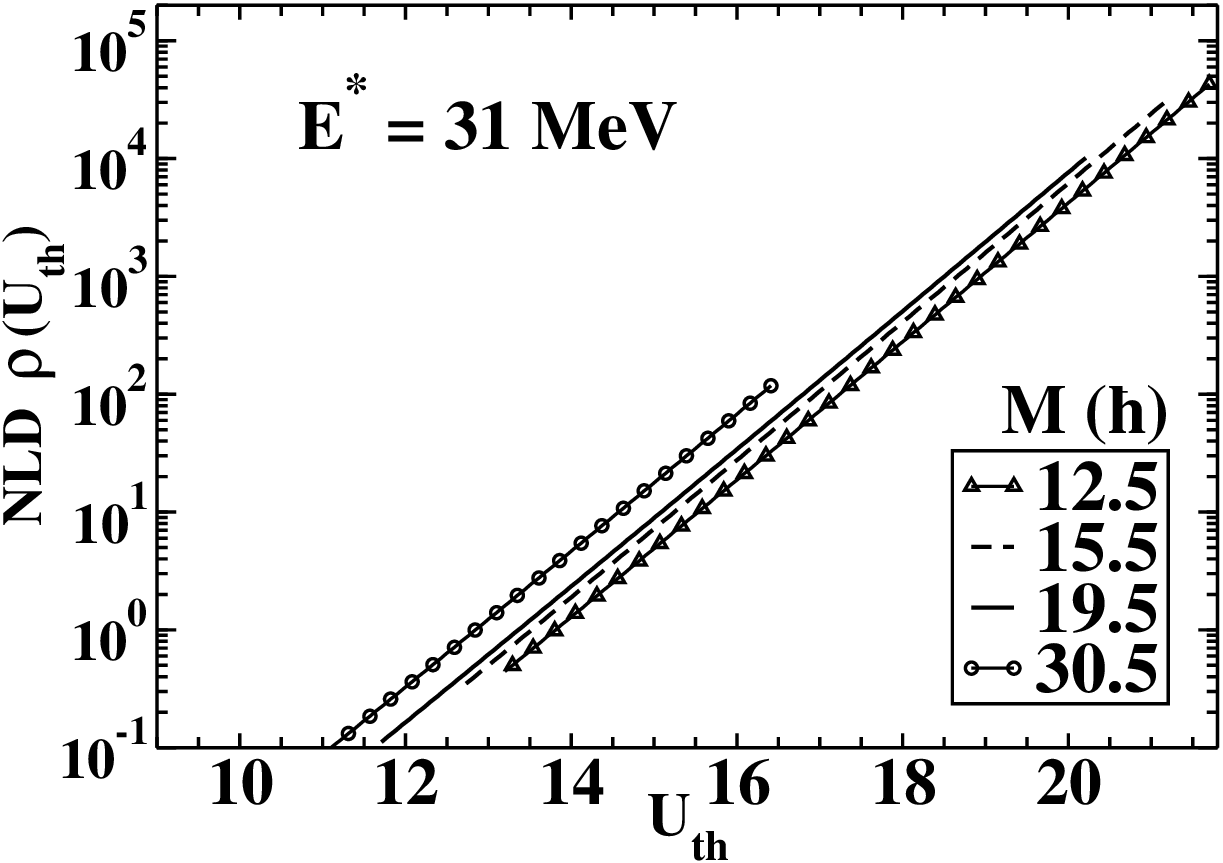}% Here is how to import EPS art
\caption{\label{fig3} Nuclear level density vs. U$_th$ for $^{119}$Sb at E$^*$ $\approx$31 MeV for M($\hbar$)$=$(a) 12.5 (b) 15.5 (c) 19.5 (d) 30.5. U$_{th}$ and NLD decrease as M increases}
\end{figure}

An estimate of NLD ($\rho$(U$_{th}$)) as a function of U$_{th}$ for E$^*$ $=$ 31 MeV and  M $=$ 12.5, 15, 15.5 and 19.5 $\hbar$ for $^{119}$Sb$^*$ is plotted in Fig. 3. The de-excitation of compound nucleus occurs by the neutron evaporation and the outgoing neutron energy E$_n$ is varied from 0$-$8 MeV. According to Eq. (1), neutron emission can occur when U$_{th}$ $>$ 0 which requires E$^*$ $>$ (S$_n$ + E$_{rot}$).  At no spin or very low spins, most of the excitation energy is available to the system and shows higher NLD. But as angular momentum increases, part of the total excitation energy is being shared with the rotational degree of freedom and as a result E$_{rot}$ increases and U$_{th}$ decreases and hence nuclear level density ($\rho$(U$_{th}$)) decreases. In case of $^{119}$Sb$^*$, $\rho$(U$_{th}$) decreases smoothly with spin due to absence of any major structural transitions. \par

\begin{figure}
\includegraphics[scale=.36]{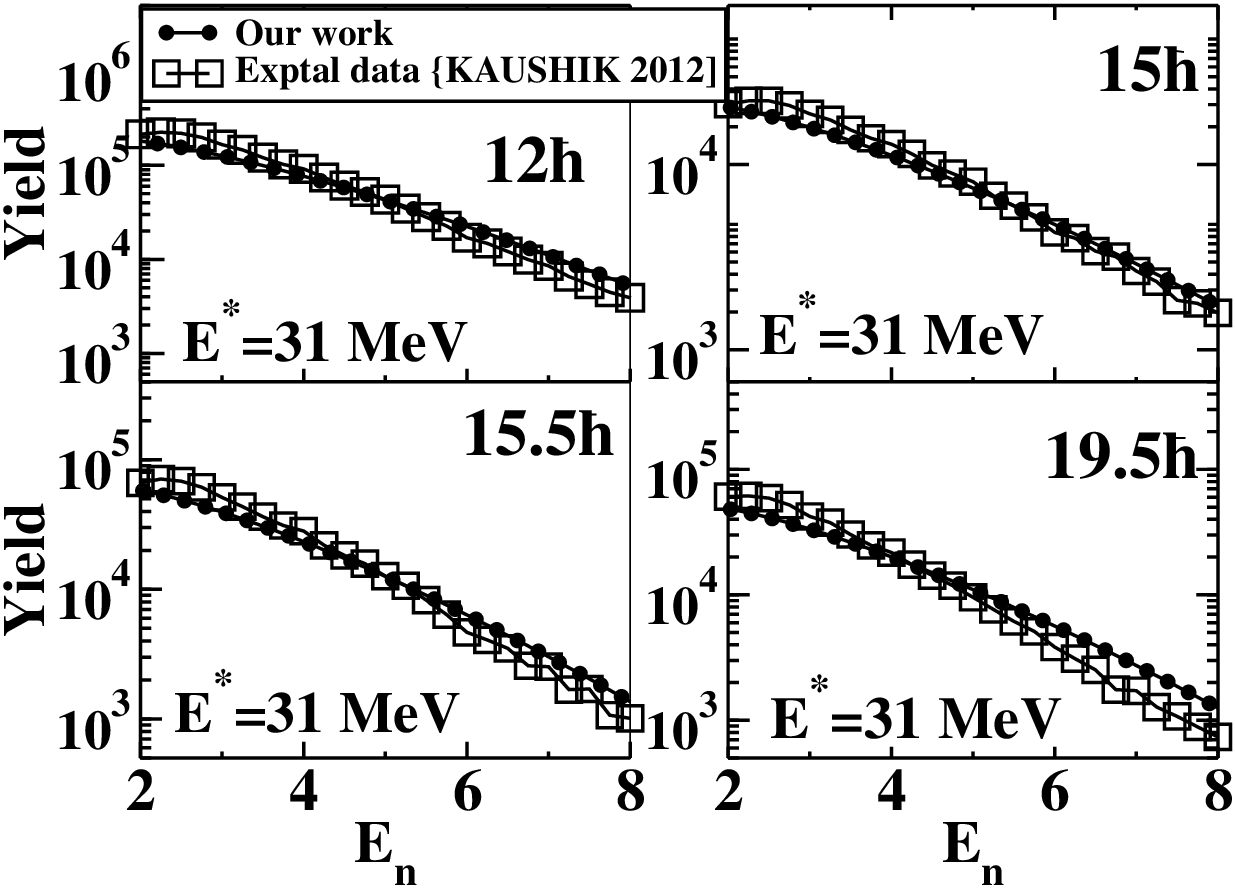}
\caption{\label{fig4} Neutron evaporation spectra for $^{119}$Sb. Available experimental data for E$^*$ $\approx$ 31 MeV is compared.}
\end{figure}
%\vspace{2.0cm}
Knowing $\rho$(U) of the residual system, one may evaluate the neutron emission probability for outgoing neutron energy E$_n$ using Eq. (1) which is compared with the experimental yield for the decay of $^{119}$Sb$^*$ plotted in Fig. 4 which show good agreement. $^{119}$Sb being close to shell closure has small deformation with no significant structural changes. The small fluctuations in deformation which are reflected in 'K' (seen in Figs. 1(a)-(c)), are not significant to show a visible influence on NLD and emission probability which appear to decrease gradually with increasing spin. \par

 However, the significance of shape transitions on NLD is evident in the case of a well deformed nuclues $^{112}$Ru in Fig. 5. $\rho$(U$_{th}$) (Fig. 5(a)) and yield (Fig. 5 (b)) decrease with increasing angular momentum in the absence of deformation phase transitions but they increase or rather jump to a much higher value for M $=$ 16$\hbar$ at which a shape transition takes place. This is a direct evidence of enhancement of level density in a deformed system undergoing a structural transition. \par

\begin{figure}
\includegraphics[scale=.38]{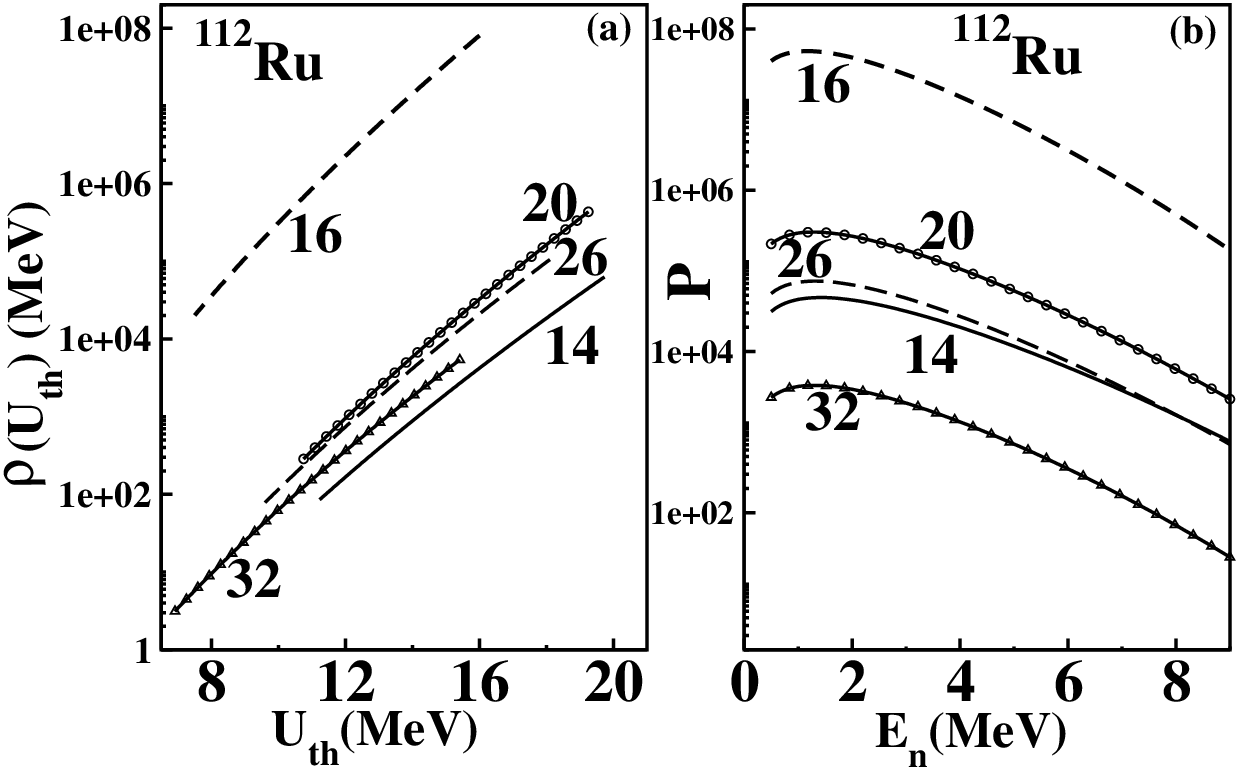}% Here is how to import EPS art
\caption{\label{fig5} Variation of (a) $\rho$(U$_{th}$) vs. U$_{th}$ and (b) Yield P vs. E$_n$ for deformed system $^{112}$Ru with  M($\hbar$) = 14, 16, 20, 26, 32. Impact of shape transition from  14$\hbar$ to 16$\hbar$ is evident}
\end{figure}

Experimentally derived Inverse level density parameter (K$=$A/a) for the decay of $^{119}$Sb$^*$ ~\cite{KAUSHIK} and $^{185}$Re$^*$ ~\cite{GOHIL} are plotted in Fig. 6.(a), (b) and (c)) along with our calculated values for M $=$11$-$21$\hbar$ and excitation energy corresponding to T values given in the respective references. The deformation and shape parameters are also plotted for the same M and T in Fig. 6 (d), (e) and (f). The points with the error bars represent the experimental data. Ref. ~\cite{KAUSHIK} predicted an overall decrease of K with spin in case of $^{119}$Sb$^*$ ~\cite{KAUSHIK} and Ref. ~\cite{GOHIL} predicted almost constant 'K' with spin in $^{185}$Re$^*$ ~\cite{GOHIL} whereas our calculated values ~\cite{MKDAE} show an overall increasing 'K' with increasing spin but show small deviations or rather decrease in 'K' at certain spin values at which the experimental 'K' also decreases in both $^{119}$Sb$^*$ and $^{185}$Re. The decrease in 'K' with spin in the case of $^{119}$Sb$^*$ is due to small fluctuations in deformation ($\beta$) (see Fig. 6(d)) and three out of four data points at M ($=$ 12.5, 15 and 15.5 $\hbar$) of  $^{119}$Sb of Ref ~\cite{KAUSHIK} show good match with our calculated 'K' values. The small dip in the value at 15 $\hbar$ agrees with a similar dip in our data which also coincides with the shape fluctuation shown in Fig. 6(d) at the same spin value. The fourth data point at M $=$19.5 $\hbar$ shows the discrepancy in the 'K' variation trend as well as the magnitude of our calculated K value. Due to the lack of experimental data for higher angular momnentum values, it is not known if the 'K' values exhibit decreasing or increasing trend for further higher angular momentum values. In case of $^{185}$Re$^*$, the constant looking 'K' plotted on a large scale in ~\cite{GOHIL} in Fig. 6 (b) and (c)) actually shows small fluctuations which are consistent with our calculated values which in turn match with the structural transitions shown in Fig. 6 (e) and (f). Here, the most important point, which is also the highlight of this work, is to note, that the decrease in 'K' values with increasing spin, in the experimental as well as in our work, is, actually due to the structural changes at those spin values which are very evidently seen in Fig. 6, which, to our knowledge, has not been reported before in any other work. The deformation and shape changes are reflected in 'K' at around/near same spin values and explain the experimental predictions ~\cite{KAUSHIK, GOHIL,PROY}. However, the available experimental data samples are too small and insufficient to establish a conclusive trend for an overall variation and hence some more exclusive measurements for a wider range of angular momentum with more data points would be helpful to get more clarity on the subject. \par

\begin{figure}
\includegraphics[scale=.42]{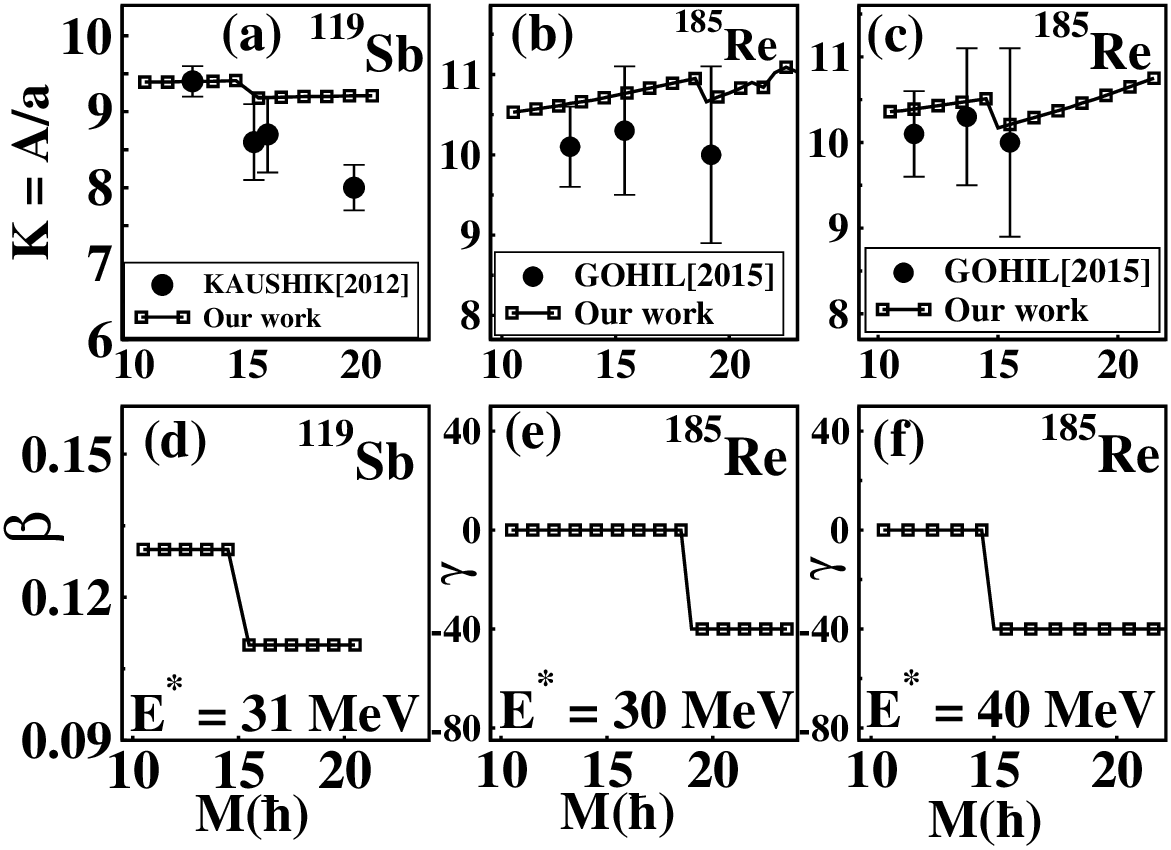}
\caption{\label{fig6} Inverse level density parameter 'K' with angular momentum M for (a) $^{119}$Sb (b) $^{185}$Re at E$^*$$\approx$ 30 MeV (c) $^{185}$Re at E$^*$$\approx$ 40 MeV. Experimental data ~\cite{KAUSHIK,GOHIL} is compared.  (d) Plot of $\beta$ vs M($\hbar$) for $^{119}$Sb (e)$\gamma$ vs M($\hbar$) for $^{185}$Re  at E$^*$$\approx$ 30 MeV (f) $\gamma$ vs. M($\hbar$) for $^{185}$Re at E$^*$$\approx$ 40 MeV}
\end{figure}

To conclude, the neutron emission spectra and NLD is estimated using the statistical theory of superfluid system using proper microscopic inputs like single particle eigen values and spin projections. Our calculations yield good results in agreement with the experimental data of neutron evaporation spectra in the decay of compound nucleus $^{119}$Sb$^*$ and prove the efficacy of our theoretical formalism. Hot and rotating compound nucleus $^{119}$Sb$^∗$ and its neighbouring N=68 isotones from $^{112}$Ru to $^{123}$Cs are investigated and a strong correlation between the spin induced deformation shape phase transitions and NLD is predicted. Our calculated inverse level density parameter 'K' increases with angular momentum for all the systems but it decreases at a deformation and a shape transition which explains the decrease in 'K' values predicted in recent experimental works. A well deformed system undergoing a sharp shape transition results in the enhancement of level density, band crossing and decrease in 'K'. These effects start diminishing while moving close to the shell closure. This shows that shell effects influence level density. Structural transition to uncommon prolate non-collective shape phase is seen for the first time in neutron rich stable nuclei $^{112}$Ru $^{113}$Rh, $^{114}$Pd, $^{115}$Ag and $^{116}$Cd at T$>$ 1 MeV which we observed earlier only in proton rich nuclei that too at very small temperatures $\approx$ 0.5-0.8 MeV. However some more experimental data samples for a wider range of spin values and masses would be helpful for further validation of theoretical results on NLD.

Financial support from  SERB, DST, Govt. of India, under WOS-A scheme is acknowledged. I thank Dr. S. Kailas for his guidance and support in the work on $^{185}$Re$^*$. I thank Dr. G. Saxena for support in peparation of manuscript and Dr. B. K. Aggarwal for useful discussion. I thank Dr. Kaushik Banerjee for discussions and providing experimental data of $^{119}$Sb.\par

\end{document}